\documentclass[useAMS,usenatbib]{mn2e}

\usepackage{epsfig}

\def\eg{{\rm e.g.}}

\def\ie{{\rm i.e.\ }}
\def\cf{{\rm cf.\ }}

\def\spose#1{\hbox to 0pt{#1\hss}}
\def\ltsimm{\mathrel{\spose{\lower 3pt\hbox{$\sim$}}
	\raise 2.0pt\hbox{$<$}}}
\def\gtsimm{\mathrel{\spose{\lower 3pt\hbox{$\sim$}}
	\raise 2.0pt\hbox{$>$}}}
\def\Mdot{\hbox{${\dot M}$}}

\def\cm{{\rm\thinspace cm}}

\def\s{{\rm\thinspace s}}

\def\g{{\rm\thinspace g}}

\def\erg{{\rm\thinspace erg}}

\def\ergpcm2ps{\hbox{${\rm\erg\cm^{-2}\s^{-1}\,}$}}
\def\pcm3{\hbox{${\rm\cm^{-3}\,}$}}
\def\gpcm3{\hbox{${\rm\g\cm^{-3}\,}$}}
\def\gpcm3ps{\hbox{${\rm\g\cm^{-3}\s^{-1}\,}$}}

\title[Numerical Simulations of Winds with Multiple Embedded Evaporating 
Clumps]
{Dynamical and Pressure Structures in Winds with Multiple Embedded
Evaporating Clumps I. 2D Numerical Simulations}

\author[J. M. Pittard, J. E. Dyson, S. A. E. G. Falle, T. W. Hartquist]
{J. M. Pittard$^{1}$\thanks{E-mail: jmp@ast.leeds.ac.uk}, 
J. E. Dyson$^{1}$, S. A. E. G. Falle$^{2}$ and T. W. Hartquist$^{1}$\\
$^{1}$School of Physics and Astronomy, The University of Leeds, 
        Woodhouse Lane, Leeds, LS2 9JT, UK\\
$^{2}$Department of Applied Mathematics, The University of Leeds, 
        Woodhouse Lane, Leeds, LS2 9JT, UK}
\begin{document}

\date{Accepted ... Received ...; in original form ...}

\pagerange{\pageref{firstpage}--\pageref{lastpage}} \pubyear{2005}

\maketitle

\label{firstpage}

\begin{abstract}
Because of its key role in feedback in star formation and galaxy
formation, we examine the nature of the interaction of a flow with
discrete sources of mass injection.  We show the results of
two-dimensional numerical simulations in which we explore a range of
configurations for the mass sources and study the effects of their
proximity on the downstream flow. The mass sources act effectively as a
single source of mass injection if they are so close together that
the ratio of their combined mass injection rate is comparable to or
exceeds the mass flux of the incident flow into the volume that they
occupy. The simulations are relevant to many diffuse sources, such as
planetary nebulae and starburst superwinds, in which a global flow
interacts with material evaporating or being ablated from the surface
of globules of cool, dense gas.
\end{abstract}


\section{Introduction}
\label{sec:intro}
Starbursts occur in regions where clumps of cool, dense, molecular
material are embedded in a far hotter and more diffuse external
medium. An understanding of the responses of such regions to winds
impacting on them is central to the investigation of feedback in star
and galaxy formation.  The same type of structure exists in, for
example, planetary nebulae, supernova remnants, H{\sc ii} regions, and the
interstellar medium of the Galactic Center.
Clumps (also referred to as clouds or globules) may either accumulate
material from the surrounding medium, and thus increase in mass, or
may lose material to the external medium, and eventually be
destroyed. In the latter case, mass loss can occur through
hydrodynamic ablation, or thermal or photoionized evaporation. The
diffuse medium is often in motion relative to these clouds, and the
nature of this interaction is of wide importance. For instance,
in the context of a starburst superwind, the observed X-ray emission
will depend on the character of this interaction.

While there have been many studies of the interaction of a dense, cold
cloud with a tenuous flow
\cite[\eg,][]{KMC1994,M1994,XS1995,GMRJ2000,LHW2001}, there have been
few calculations of the interaction between multiple clouds and
a flow \cite[\eg,][]{JJN1996,PFB2002}. While the latter have recently
been supplemented by some wonderful laser experiments (\cite{P2004};
see also \cite{KBPB2003}), our understanding of such interactions is
still developing.

A limitation of previous models is that the clouds have been modelled
as single phase entities, and the density contrast between the
cloud(s) and the flow has typically been taken to be of the order of
$10^{2}$ (for numerical reasons). The simulated clouds then have such
short lifetimes that they are unable to significantly \lq
massload\rq~the flow. In reality, the astrophysical clouds which are
of interest are cold and molecular, and have much larger density
contrasts. In this case, the time required to destroy the clouds is
much longer than other relevant time-scales and the rate of mass-loss
from the cloud can be assumed constant. In this limit the mass-loss
may significantly massload the flow, with the injected material
occupying a cone with a sizeable opening angle if the wind is
hypersonic, or being confined to a long, thin tail when the wind is
transonic \citep{DHB1993,FCPDH2002}. A short review of tail formation
by this, and other processes, is given in \cite{D2003}.

The nature of the interaction of a flow with a large group of clouds
can differ substantially from that which occurs with a single cloud.
If the diffuse medium surrounding embedded molecular clouds is flowing
supersonically, then the clouds are likely to be destroyed by
hydrodynamical ablation.  Mass injection into the flow due to the
destruction of the clouds may sometimes greatly enhance the thermal pressure of
a flow at the expense of the flow's ram pressure. The properties of
the flow may then be much more conducive to the survival of clouds
further downstream, and if the flow is slowed and pressurized enough,
it may even induce their collapse and hence trigger new star
formation. This process could be a central mechanism for feedback in the
interstellar medium (\eg, in starburst regions).

\begin{figure}
\begin{center}
\psfig{figure=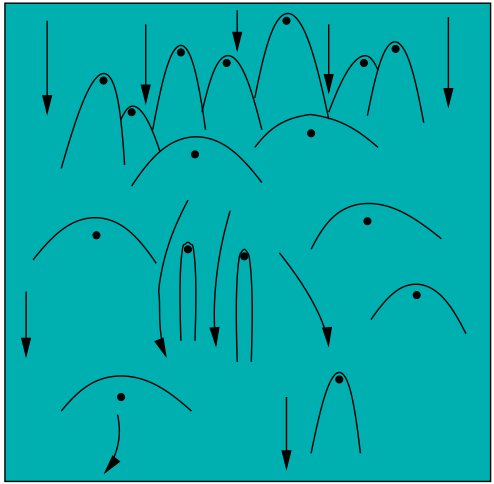,width=6.5cm}
\end{center}
\caption[]{A schematic diagram which illustrate the main features which
we might expect when a supersonic, tenuous flow encounters a 
region containing many discrete clouds. The clouds are not drawn to scale.}
\label{fig:ml_schematic}
\end{figure}

The main features which we might expect to see in the interaction
between multiple clouds and a tenuous supersonic flow are shown
schematically in Fig.~\ref{fig:ml_schematic}. Individual bowshocks
form around those clouds furthest upstream and merge at some point
downstream. While much of the material in the flow is likely to remain
supersonic in this region, further encounters with clouds and the
creation of additional bowshocks means that the flow will gradually
slow, pressurize, and become subsonic. New star formation may occur in
this part of the flow. Other clouds may continue to lose mass, but
since they interact with a subsonic flow their injected mass will be
confined to long thin tails. Such tails have only a small
cross-section to the oncoming flow, and do not greatly impede it. The
flow may therefore accelerate between the clouds as this region
effectively becomes more porous, and may become supersonic again. If
more clouds are encountered further downstream the whole scenario may
repeat.  One issue is whether such a system would reach a steady
state, or whether it would flicker due to mass pickup by the diffuse
flows being less effective when the flow is transonic compared to when it is
supersonic.

In this paper we use hydrodynamical calculations to investigate the
nature of the interaction between a tenuous flow and a number of
mass sources. In Section~\ref{sec:model} we describe the problem and
the basic assumptions used. The numerical results for single to
multiple mass sources interacting with both a hypersonic and transonic
wind are presented in Section~\ref{sec:results}. Section~\ref{sec:conclusions} 
contains our conclusions and ideas for future work.


\section{The Model}
\label{sec:model}
As our investigation is still in its early stages, in this study we
concern ourselves with only the general properties of the interaction
of a wind with injected material. We do not, therefore, attempt to
model the detail of {\em how} material is injected.  Instead, we
follow the approach in \cite{FCPDH2002} by assuming a uniform rate of mass
injection within a given radius.  This is simple to implement, but has the
disadvantage that the flow from the injection region is isotropic. The
mass loss from a cloud will not be isotropic if it is caused by
hydrodynamic ablation by an incident wind, or by photoevaporation by
radiation from a nearby star, but we show in
Section~\ref{sec:1-2hyper} that the effect of asymmetrical mass loss
is small at large distances from the cloud.  Therefore, the actual
details of the mass injection process are relatively unimportant.  We
also emphasize that the boundary of this injection region {\em is not}
meant to be the boundary of a cloud. In fact, the cloud could be much
smaller.
 
In this paper we are concerned with the interaction of a flow incident
on several mass sources. Three dimensional calculations are required
if the sources are spherical. To reduce the
computational cost we restrict ourselves to two dimensional
simulations where the sources are cylindrical. While we expect some 
differences between calculations performed in 2D and 3D, at this stage
we can still gain important insight from less computationally demanding 
2D simulations. 

We investigate the simplest case in which the incident wind behaves 
adiabatically and the injected gas remains isothermal. This is
motivated by the fact that if mass injection is due to
photoevaporation, the injected temperature will be $\sim 10^{4}$~K, 
which is comparable to that of a wind whose temperature is determined
by photoionization - the wind, however, can be shocked to much higher 
temperatures, which is the reason why we can see tails from mass sources
in many astrophysical media\footnote{If the Mach number of the wind were
lower, the density contrast between the injected material and the wind
would be reduced, in which case the described dichotomy is a
poorer approximation.}. To ensure the above behaviour, we
use an advected scalar, $\alpha$, which is unity in the injected gas
and zero in the ambient gas. The source term in the energy equation is 
then
\begin{equation}
\label{eq:esource}
K \alpha \rho (T_{\rm 0}-T),
\end{equation}
where $\rho$ and $T$ are the local mass density and temperature, and
where $K$ is large enough that the temperature always remains close to
the equilibrium temperature, $T_{\rm 0}$, in the injected gas.  Inside
the injection region we add an extra energy source so that the gas is
injected with temperature unity (see \cite{FCPDH2002} for further
details).

The calculations reported in this paper use Cobra, a 2nd order accurate
code with adaptive mesh refinement (AMR). Cobra uses a hierarchy of grids
$G^{0}...G^{N}$ such that the mesh spacing on grid $G^{n}$ is 
$\Delta x_{0}/2^{n}$. Grids $G^{0}$ and $G^{1}$ cover the whole domain, but
the finer grids only exist where they are needed. The solution at each
position is calculated on all grids that exist there, and the difference
between these solutions is used to control refinement. In order to
ensure Courant number matching at the boundaries between coarse and fine
grids, the time step on grid $G^{n}$ is $\Delta t_{0}/2^{n}$ where 
$\Delta t_{0}$ is the time step on $G^{0}$. Such a hierarchical
grid structure not only improves the efficiency by confining the fine grids
to where they are needed, but it also makes it possible to use a full
approximation multigrid algorithm to accelerate the convergence to the
steady state \cite[see, \eg,][]{B1977}. Further details of Cobra can be 
found in \cite{FG1993}. 

\section{Results}
\label{sec:results}
\subsection{Single and Twin Mass Sources}
\label{sec:1-2sources}
Our first investigations are of one and two mass sources interacting
with an ambient flow.
So that the ambient wind does not affect the flow in the injection region,
we must ensure that the ram pressure of the injected material at the
boundary of the injection region is larger than that in the wind,
\ie we require
\begin{equation}
\label{eq:mass_inj}
a^{2} \rho_{\rm s} = \frac{a r_{\rm c} Q}{3} \geq \rho_{\rm w}v_{\rm w}^2,
\end{equation}
where $r_{\rm c}$ is the radius of the injection region, $a$ is the 
flow speed of injected material at $r=r_{\rm c}$ and
$\rho_{\rm s}$ its density at this point, $Q$ is the mass injection rate
per unit volume within $r=r_{\rm c}$, and $\rho_{\rm w}$ and $v_{\rm w}$ are
the wind density and velocity (\cf Equation~5 in \cite{FCPDH2002}).
We again emphasize that the cloud radius could actually be much smaller 
than $r_{\rm c}$.

\subsubsection{Hypersonic wind}
\label{sec:1-2hyper}
We choose units such that $r_{\rm c}=1$, $a=1$, $\rho_{\rm w}=1$, and
set $v_{\rm w}=20$ and $T_{\rm 0}=1$, which correspond to an external
isothermal Mach number of 20. Equation~\ref{eq:mass_inj} then implies
$Q \geq 800$ in order to prevent the wind from penetrating the mass
injection region, although \cite{FCPDH2002} note that in the
supersonic case a better estimate is obtained by replacing $\rho_{\rm
w} v_{\rm w}^2$ by the pressure behind a stationary normal shock in
the wind. This gives
\begin{equation}
\label{eq:mass_inj2}
Q \geq \frac{4}{\gamma+1} \frac{\rho_{\rm w}v_{\rm w}^2}{a r_{\rm g}} = 600
\end{equation}
for $\gamma=5/3$. We set $Q=700$ in order to ensure that the interaction
occurs slightly outside the mass-injection region. As in \cite{FCPDH2002}, we
set the coefficient $K$ to 50, which is large enough to ensure that the
temperature in the injected gas stays close to unity.

The computational domain is $0 \le x \le 50$, $-50 \le y \le 20$ with
the mass injection region initially centered at the origin. 5 grid
levels $G^{0}...G^{4}$ were used, with $G^{0}$ being $50\times70$ and
$G^{4}$ $800\times1120$. The diameter of the mass injection region is
32 cells on the $G^{4}$ grid. Artificial dissipation is added to the
simulations in order to eliminate the \lq carbuncle effect\rq\footnote{This
is an unphysical distortion of a shock front that is partially aligned 
with the grid \citep{Q1994}. It can be cured by adding a small amount of 
artificial viscosity to the Riemann solver \cite[\eg,][]{FKJ1998}. This
artificial dissipation is quite separate from the turbulence 
model noted in Section~\ref{sec:1-2trans}, and its influence is confined 
to the grid scale.}~and to damp
Kelvin-Helmholtz instabilities due to velocity shear at the contact
discontinuity. This allows a steady-state solution.

Fig.~\ref{fig:hypv1} shows the density, pressure, and y-velocity,
together with the regions occupied by the finest grid. The wind is
decelerated by a bow shock, which stands off from the center of the
mass source by a distance of approximately 8 units.  A contact
discontinuity separates the wind from the injected material.  As noted
in \cite{FCPDH2002} for the case of a spherical mass source, the injected
material is not confined to a long thin tail. Instead the contact
discontinuity has a half-width of $\approx 20$ at $y=-50$, which is
much wider than the mass source. It is also much wider than the
equivalent width observed at the same distance downstream from a
spherical mass source, due to the reduced divergence in
the 2D simulation. It appears that the contact discontinuity does not
reach an asymptotic off-axis distance far downstream, but rather tends
towards a finite opening angle (this can be further discerned in
Fig.~\ref{fig:multiv4}). Though not shown here, the opening angle is
dependent on the Mach number of the wind - the injected material is
more confined at lower Mach numbers, though the opening angle of the
bowshock is greater. A reverse shock surrounds the mass
source, and delineates the position at which the isotropic injected
material feels the presence of the ambient wind.

\begin{figure*}
\begin{center}
\end{center}
\caption[]{Density, pressure, y-velocity, and $G^{4}$ for the
simulation of a hypersonic flow interacting with a single
cylindrical mass source.}
\label{fig:hypv1}
\end{figure*}

In Fig.~\ref{fig:noniso} we show density plots from simulations
where the density, velocity, and pressure of the injected material is
specified around the edge of the injection region. This approach allows
us to investigate the effect of non-isotropic mass injection on the 
interaction with the ambient wind. In the simulation shown in the left
panel of Fig.~\ref{fig:noniso} we set $\rho_{\rm s}=350$ and $a=1$, and
keep the other parameters as before. The total mass-injection rate is the same
as that in Fig.~\ref{fig:hypv1}, though the energy injection is slightly
different. The latter means that there are slight differences between 
the density plots in Figs.~\ref{fig:hypv1} and~\ref{fig:noniso}.
In the middle panel of Fig.~\ref{fig:noniso} we keep the same
overall mass injection rate, but vary the density at the edge of the 
injection region according to the prescription
\begin{equation}
\label{eq:rho_noniso}
\rho_{\rm s} = \rho_{0} (1 - \Omega^{2}\sin^{2}\theta),
\end{equation}
where $\rho_{0}$ is a normalization factor, $\theta$ is the angle
between the radial vector from the injection region and the ambient
flow ($\theta=\pi$ on the upstream surface, and $\theta=0$ on the 
downstream surface), and $\Omega$ sets the degree of anisotropy.
We set $\Omega=0.9$, so that the mass injection rate at the upstream
and downstream surfaces is $\approx 5 \times$ that at $\theta=\pi/2$.
While there are differences in the morphology of the interaction close
to the injection region (\eg, the bow shock stands further off,
and the shape of the reverse shock reflects the latitudinal variation
of the mass and energy injection), the large scale features are
remarkably unchanged. In the right panel of Fig.~\ref{fig:noniso} 
the injection rate is highest at the upstream surface and declines 
smoothly towards the downstream surface, according to the prescription
\begin{equation}
\label{eq:rho_noniso2}
\rho_{\rm s} = \rho_{0} [1 + \Omega\sin(\theta+\pi/2)].
\end{equation}
This prescription may be expected to be more representative of reality. 
With $\Omega=0.7$, the mass injection rate at the upstream surface is 
$\approx 5 \times$ that at the downstream surface. Again we see that
the large scale features are very similar, though as expected the 
bowshock stands slightly further off than in the other two cases.

\begin{figure*}
\begin{center}
\end{center}
\caption[]{Density plots for the simulation of a hypersonic flow
interacting with a single cylindrical mass source, where the injected
material is isotropic (left) and non-isotropic (middle and right). The
large-scale structure is similar in all cases.}
\label{fig:noniso}
\end{figure*}

\begin{figure*}
\begin{center}
\end{center}
\caption[]{Density plots for the
simulation of a hypersonic flow interacting with two
cylindrical mass sources as a function of their separation, which
increases from 12 units (top left), to 24 units (top right), to
48 units (bottom left), to 96 units (bottom right).}
\label{fig:hypv2}
\end{figure*}

In Fig.~\ref{fig:hypv2} we show the density from calculations where
the mass source is moved off-axis, which simulates the interaction
of a wind with 2 identical sources. The separation between the two
sources increases from 12, to 24, to 48, to 96 units. In the top left 
panel of Fig.~\ref{fig:hypv2} the two sources are surrounded by a global
contact discontinuity, and not individual ones. The bow shock is
located further upstream compared to that in Fig.~\ref{fig:hypv1}, and
is also global in the sense that it envelops the two mass sources, as
opposed to individual shocks surrounding each source.  
When the mass sources are still fairly close together, the flow
between them is at a higher pressure than the corresponding flow
around the outside edge of the interaction. This causes the flow
around each mass source to angle away from each other, and is easily
identified by the tilt of the reverse shock around each source
relative to the flow of the ambient wind.

As the mass sources are separated, first the global contact
discontinuity splits into individual contacts around each source, and
then the global bow shock separates into individual bow shocks.
The reverse shock around each mass source
is once again aligned with the oncoming wind, and the tail of
injected material is initially symmetrical and unaffected by the presence of
the other mass source. However, some interaction occurs further 
downstream where the bowshocks interact. At this point a reflected 
shock is formed, and further downstream this deflects the contact 
discontinuity and the tail of injected material away from the axis 
of symmetry. The simulations shown in Fig.~\ref{fig:hypv2}
used 4 grid levels $G^{0}...G^{3}$. The diameter of the mass injection region
is 16 cells on the $G^{3}$ grid. For the models with separations of
12, 24, and 48 units, the $G^{3}$ grid is $400 \times 560$, and the
computational domain is $0 \le x \le 50$, $-50 \le y \le 20$.
The model with a separation of 96 units has a $G^{3}$ grid of 
$768 \times 960$, and a computational domain which encompasses 
$0 \le x \le 96$, $-100 \le y \le 20$.

\subsubsection{Transonic wind}
\label{sec:1-2trans}
In many situations it is usual for a wind to encounter a large number
of mass sources. In such circumstances the Mach number of the wind is
driven towards unity \citep{HDPS1986}, and it is therefore reasonable
to look at the interaction of mass sources with a transonic wind. In
this case we set $r_{\rm c}=1$, $a=1$, and $T_{\rm 0}=1$ as before but
now $\rho_{\rm w} = 10^{-3}$, $v_{\rm w} = 40.825$, giving a Mach
number of unity in the undisturbed wind. Equation~\ref{eq:mass_inj}
gives $Q \geq 10/3$, which is a good estimate for this case since
there is no shock upstream of the injection region. We therefore set
$Q = 10/3$, and the coefficient $K$ to $10^{2}$ \citep[\cf][]{FCPDH2002}.

\begin{figure*}
\begin{center}
\end{center}
\caption[]{The effect of the turbulence model on a calculation where a
cylindrical source of mass injection interacts with a transonic wind. In the
left panel we solve only the Euler equations. In the right panel we
add a turbulence model where velocity shear is converted into a scalar
which represents the turbulent energy density, and an additional scalar
controls the dissipation rate. The resulting solution is an approximation
to the mean flow, and the mixing of the injected gas with the original flow
is modelled by the diffusive terms in the equations. In the left panel, the
instabilities in the tail occur further upstream if the numerical resolution 
of the calculation is increased, since the numerical viscosity is reduced. 
In contrast, the turbulence model in the simulation in the right panel prevents
resolution dependent instabilities.}
\label{fig:turb}
\end{figure*}

Simulations with a transonic wind pose additional difficulties
compared to the hypersonic wind case. First, in order for the
boundaries to have no effect on the solution, the computational domain
has to be very large, particularly since our 2D simulations have
reduced divergence compared to the case of spherical mass injection
regions. To satisfy this condition we find that we need $0 \le x \le
320$, $-416 \le y \le 256$, which would have made the calculation
extremely expensive if an AMR code were not employed.  Second, the
velocity shear between the injected material and the wind is so
extreme in the transonic case that the wind flow separates and
produces a turbulent wake downstream of the interaction region.
Calculations based on the Euler equations cannot adequately describe
such turbulence, and the only viable option is to use a turbulence
model.  While there are many possibilities, we use a simple
$k-\epsilon$ model as used in \cite{FCPDH2002}. 

The purpose of the subgrid turbulence model is to emulate a high 
Reynolds number flow.  It does so by including equations for the turbulent 
energy density and dissipation rate and using these to calculate viscous 
and diffusive terms. The resulting solution should be an approximation to 
the mean flow. The turbulent mixing of the injected gas with the original 
flow due to shear instabilities is modelled by the diffusive terms in the 
equations.  Since the turbulent viscosity computed from the subgrid model 
depends upon the local solution, it is not very meaningful to talk about 
an effective Reynolds number. For example, the turbulent viscosity is 
largest in shear layers and essentially vanishes in regions with little 
shear. The model has been calibrated by comparing the computed 
growth of shear layers with experiments \citep{DW1983}. 
The model also assumes that the real 
Reynolds number is very large, which is the case in astrophysical flows, 
and that the turbulence is fully developed. 
Although not entirely satisfactory, such a model gives a much more 
realistic result than simply using grid viscosity since in that case the 
size of the shear instabilities are determined by the numerical 
resolution. Further details can be found in
\cite{F1994}. The effect of the turbulence model is illustrated in
Fig.~\ref{fig:turb}.  


We use 9 levels of grid refinement for our transonic simulations, 
$G^{0}...G^{8}$, with $G^{0}$
being $10 \times 21$ and $G^{8}$ $2560 \times 5376$. The diameter of the
mass injection region is 16 cells on the $G^{8}$ grid. The interaction
is very different from the hypersonic case, as can be seen from
Fig.~\ref{fig:transv1}. Instead of a bow shock there is a bow wave
upstream of the mass source, whose amplitude falls off as $1/r$. A
very weak tail shock occurs in the wind downstream of the mass source,
and this is aligned much more parallel to the ambient flow than when
the mass source is spherical \cite[\cf][]{FCPDH2002}.  The injected material
remains in rough pressure equilibrium with the wind, and eventually
becomes confined to a tail whose width is of about the same order as
the injection region.

In Fig.~\ref{fig:transv2} we show the density, pressure, and
y-velocity from a calculation where the mass source is moved off-axis,
as we again simulate the interaction between a wind and 2 identical
sources.  The separation between the two sources is 48 units. We
immediately see that the tail produced behind each mass source is
strongly curved towards the other, producing a narrow channel between
the tails. However, it is clear that the tails are initially tilted
away from each other, as shown by the position of the reverse shock
around the mass source. The widening of the channel between the two
mass sources, which is enhanced by the curvature of the inner side of
the contact discontinuity, causes the pressure within the channel to
drop via the Bernoulli effect as the wind accelerates and becomes
supersonic.  The fall in pressure relative to that on the outside edge
of the tail results in the tail curving towards the axis of symmetry,
and the channel is then closed off. A shock in the channel just above
this point slows the accelerated wind. 
There is no sign of any tail
shock in the downstream wind on the outer side of the tail.

\begin{figure*}
\begin{center}
\end{center}
\caption[]{Density, pressure, and y-velocity for the
simulation of a transonic flow interacting with a single
cylindrical mass source.}
\label{fig:transv1}
\end{figure*}

\begin{figure*}
\begin{center}
\end{center}
\caption[]{Density, pressure, and y-velocity for
a transonic flow interacting with two
cylindrical mass sources separated by 48 units.}
\label{fig:transv2}
\end{figure*}

Further simulations reveal that the curvature of the tails
towards each other decrease
when the separation between the mass sources is increased.
At large enough separations there is no interaction between
the mass sources and the tails remain perfectly straight.

 
\subsection{Multiple Sources in a Hypersonic Wind}
\label{sec:multiple}
In this section we investigate the interaction between a hypersonic flow
and a group of 5 mass sources (10 with the imposed symmetry), and use the
turbulence model noted in Section~\ref{sec:1-2trans} for these calculations.

In the first simulation the sources are randomly distributed within a
circular region of diameter 160 units. This diameter is increased in
subsequent simulations, in order to explore differences between an
interaction which is dominated by the mass injection to one which is
dominated by the wind, though the relative positions of the sources
remain the same. We define the parameter $\chi = \Mdot_{\rm
c}/\Mdot_{\rm w}$, where $\Mdot_{\rm c}$ is the combined mass
injection rate of the sources and $\Mdot_{\rm w}$ is the mass flux in
the wind through a suitably chosen region.  When we change the size of
the region in which the mass sources are distributed, we alter the
value of $\chi$ since $\rho_{\rm w}$ and $v_{\rm w}$ are kept
constant.  The mass injection regions have diameters of 8 cells on the
finest grid in each of the models presented in this section.

The positions of the 5 sources in our first simulation are $(x,y)=
(33.72,-57.12)$, (48.69,-13.70), (39.07,2.54), (24.68,55.46), and
(7.89,-14.51). The best estimate for $\Mdot_{\rm w}$ is obtained from
the evaluation of the flow rate through a region bounded by the
symmetry axis and the source region whose $x$-coordinate is furthest
from this. $D$ is set to the value of this $x$-coordinate.  Thus
$\Mdot_{\rm w} = \rho_{\rm w} v_{\rm w} D = 20 \times 1 \times 48.69 =
973.8$. Because we are adding mass and energy to the wind, we use a
slightly higher value of $Q$ in these simulations to ensure that the
flow does not enter any of our source regions ($Q=880$).  Therefore,
$\chi = 5 \pi \times 880/973.8 \approx 14$ and we expect the
interaction to be injection-dominated. We use 5 grid levels,
$G^{0}...G^{4}$ with $G^{0}$ being $40 \times 60$ and $G^{4}$ $640
\times 960$. The computational domain is $0 \le x \le 160$, 
$-160 \le y \le 80$. 

In Fig.~\ref{fig:multiv1} we show the density, pressure, y-velocity,
and the advected scalar of this model. A global bow shock exists
around the group of mass sources and the region between the sources is
filled with high pressure, low Mach number gas. The shape of the
global bow shock is to some extent determined by the positions of the
individual sources, and has an opening angle which is much wider than
that obtained when there are only one or two sources (\cf
Section~\ref{sec:1-2hyper}).  Downstream of the mass sources the
injected material is accelerated by a pressure gradient in a manner
similar to that of a superwind. The shape and position of the reverse
shock around each mass source is dependent on the local flow
conditions, and in some cases is almost spherical. Close up images of
some of the injection regions are shown in Fig.~\ref{fig:multiv2}. The
bottom right panel of Fig.~\ref{fig:multiv1} reveals that the group of
mass sources is largely impervious to the oncoming wind. The mass
source which is furthest upstream is somewhat akin to a continental
divide in the sense that it splits the wind flow to the left or
right. Since we impose symmetry at $x=0$ a high pressure region of
shocked ambient wind is formed - wind in this region is able to
percolate through the region of mass sources, and is the only part of
the oncoming flow which is able to do so.

\begin{figure*}
\begin{center}
\end{center}
\caption[]{Density, pressure, y-velocity, and advected scalar for the
interaction of a hypersonic flow with multiple 
cylindrical mass sources.}
\label{fig:multiv1}
\end{figure*}

\begin{figure*}
\begin{center}
\end{center}
\caption[]{Density images of the flow in the vicinity of three of the 
injection regions in the
simulation shown in Fig.~\ref{fig:multiv1}.}
\label{fig:multiv2}
\end{figure*}



If the distances between the mass sources are increased by a factor of
4, $\chi$ is reduced to $\approx 3.5$. The interaction between the
wind and the injected material is still dominated by the mass sources,
as shown in Figs.~\ref{fig:multiv3} and~\ref{fig:multiv3b}, but to a
lesser extent than the simulation shown in Fig.~\ref{fig:multiv1}.
The mass source furthest upstream again acts like a continental
divide.  However, the flow in the vicinity of this source is identical
to the single source case, being unaltered by the complexities of the
interaction further downstream. Specifically, the reverse shock around
the mass source is not tilted or compressed. Where the interaction with $\chi
\approx 14$ is characterized by a large region of subsonic flow, the
interaction with $\chi \approx 3.5$ has a complex structure with
multiple shocks, as is readily apparent in
Fig.~\ref{fig:multiv3}. This is due to the fact that the distances
between the mass sources are such that pressure gradients in the
vicinity of each source accelerate the flow to supersonic velocities
before additional mass sources are encountered further downstream. 
In addition to a region near the axis of
symmetry, the advected scalar reveals that the wind is able to force its way
between the mass sources in two distinct streams, which become
diluted by injected material along their length. In
this simulation we used 7 grid levels, $G^{0}...G^{6}$ with $G^{0}$
being $40 \times 60$ and $G^{6}$ $2560 \times 3840$, spanning
a computational domain of $0 \le x \le 640$, $-640 \le y \le 320$.
 
\begin{figure*}
\begin{center}
\end{center}
\caption[]{The density (left panel) and advected scalar (right panel) 
in the interaction of a hypersonic flow with multiple 
cylindrical mass sources when $\chi \approx 3.5$.}
\label{fig:multiv3}
\end{figure*}

\begin{figure*}
\begin{center}
\end{center}
\caption[]{Enlargements of the regions around the 5 mass 
sources in the simulation shown in Fig.~\ref{fig:multiv3}.}
\label{fig:multiv3b}
\end{figure*}

An increase in the separation of the mass sources by another factor of
4 reduces $\chi$ to approximately unity. We now expect the wind to
begin to force its way through the region of mass sources, and this is
demonstrated in Fig.~\ref{fig:multiv4}, a simulation making use of 9
grid levels, $G^{0}...G^{8}$ with $G^{0}$ being $40 \times 60$ and
$G^{8}$ $10240 \times 15360$ (157 million cells equivalent). The
computational domain spans the region $0 \le x \le 2560$, $-2560 \le y
\le 1280$.  The distances between each mass source are now so great
that there is very little interaction between them, the main
difference being that the sources which are furthest downstream are
interacting with a supersonic flow whose properties have been somewhat
modified by the action of the sources further upstream.

\begin{figure*}
\begin{center}
\end{center}
\caption[]{The density (left panel) and advected scalar (middle panel)
in the interaction of a hypersonic flow with multiple cylindrical mass
sources when $\chi \approx 1$.  An enlargement of the flow around two
of the mass sources is displayed in the right panel.}
\label{fig:multiv4}
\end{figure*}

\begin{figure*}
\begin{center}
\psfig{figure=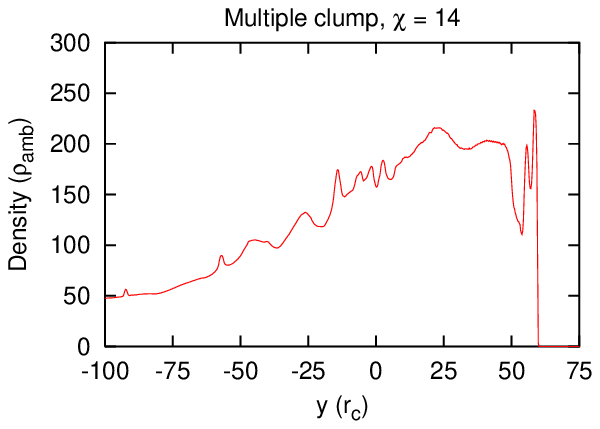,width=8.0cm}
\psfig{figure=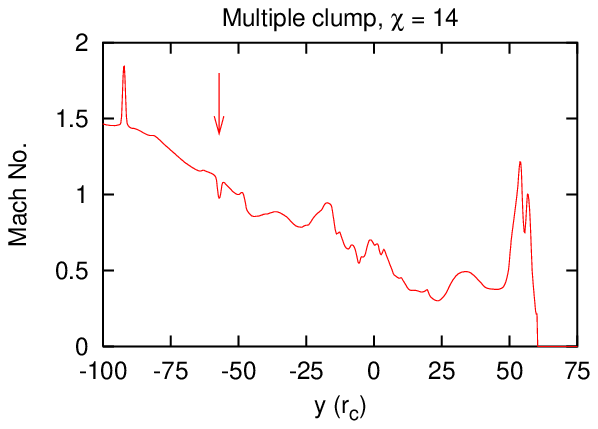,width=8.0cm}
\psfig{figure=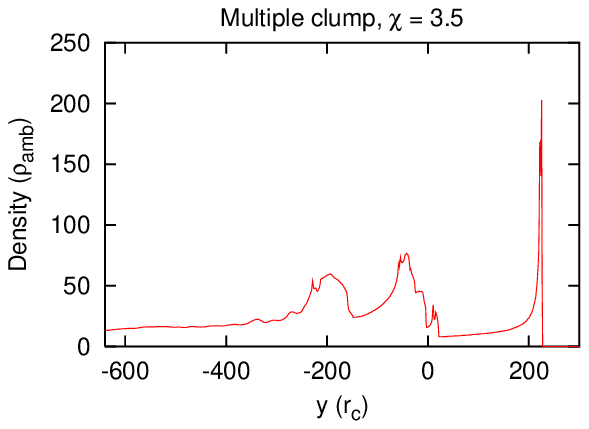,width=8.0cm}
\psfig{figure=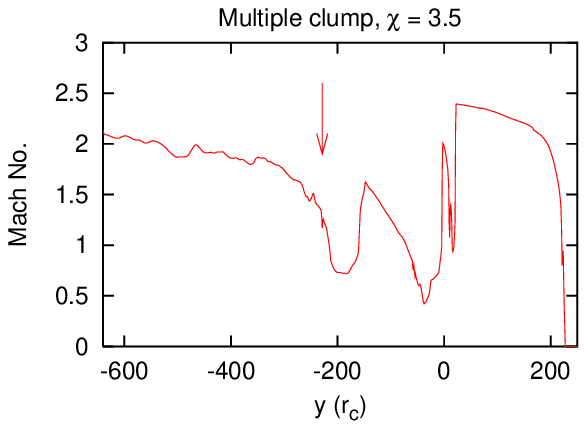,width=8.0cm}
\psfig{figure=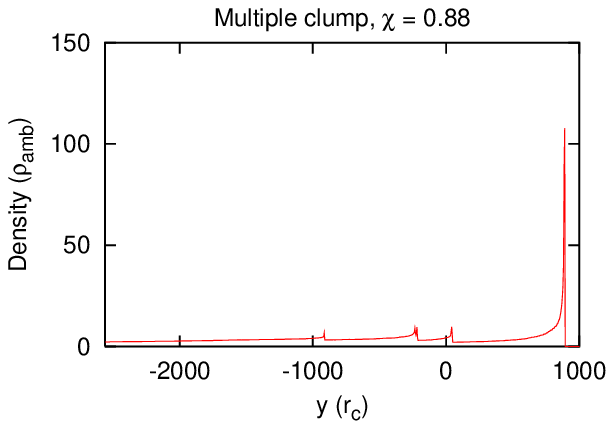,width=8.0cm}
\psfig{figure=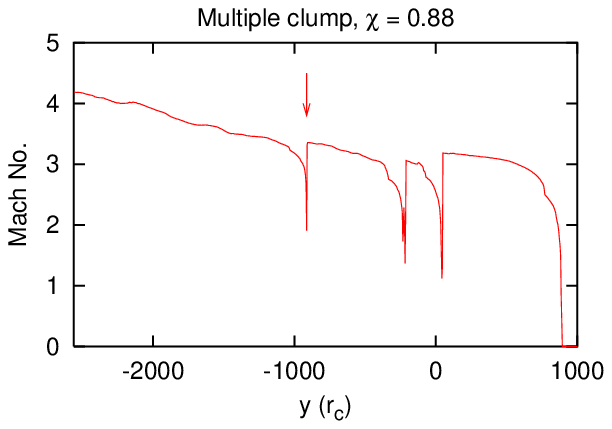,width=8.0cm}
\end{center}
\caption[]{The density and Mach number averaged across the interaction
region as a function of downstream distance for a supersonic wind
impinging on a region containing many mass sources. The parameter
$\chi$ describes the ability of the wind to penetrate through the
region containing the mass sources.  The arrows marked on the Mach
number plots illustrate the $y$-coordinate of the mass source that is
furthest downstream.}
\label{fig:multiv5}
\end{figure*}

The density and Mach number averaged across $x$ in the disturbed flow
as a function of $y$ for each of the three simulations discussed in
this section are shown in Fig.~\ref{fig:multiv5}. The Mach number is
mass averaged and not volume averaged. We see that when the rate of
mass injection dominates the mass flux of the wind (\ie $\chi \gg 1$)
the average density across the width of the interaction region is very
high, being of order 200 times the ambient density of the wind over a
large volume of the region that contains the mass sources. The
corresponding Mach number in this case is typically 0.5. The density
decreases and the Mach number increases downstream of the region
containing the mass sources, and the flow passes through a sonic point
at roughly the same $y$-coordinate as that of the most downstream mass
source. The density and Mach number profiles are fairly smooth, though
there are some features with small spatial scales.

As $\chi$ decreases, the mass sources have a much more
localized effect on the flow. When $\chi=3.5$, we see that the flow
between the most upstream mass source (at $y \approx 222$) and its
nearest companion (at $y \approx 10$) becomes supersonic, and thus
knows nothing of the presence of this second source prior to the
bowshock around it\footnote{This is in marked contrast to the
simulation with $\chi=14$ where a high pressure region moves upstream
and compresses the reverse shock around the mass source furthest upstream.}. 
The density in this part of the flow thus
significantly decreases from its peak post-shock value around the
first mass source ($\rho \approx 200 \rho_{\rm amb}$) until its
encounter with the second mass source ($\rho \approx 8 \rho_{\rm amb}$
just ahead of the bowshock). There are a few places in the
interaction region where the flow averaged over the width of the
interaction is subsonic, and these are associated with local maxima in the
corresponding density profiles. Once again, the overall flow passes
through a sonic point close to the $y$-coordinate of the most downstream
mass source. Compared to the simulation with $\chi=14$, the density
enhancement of the mass-loaded flow is much reduced, and the averaged flow
is able to maintain a higher Mach number.
The trends which we have noted above continue as $\chi$ is made yet 
smaller. When $\chi=0.88$ the individual mass sources appear to
act only as localized perturbations on the flow.

\begin{figure*}
\begin{center}
\psfig{figure=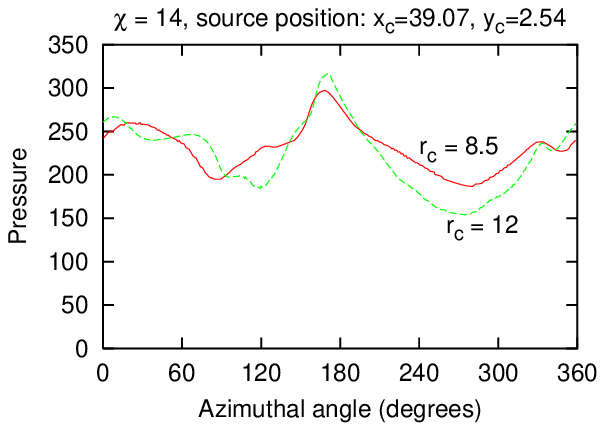,width=8.0cm}
\psfig{figure=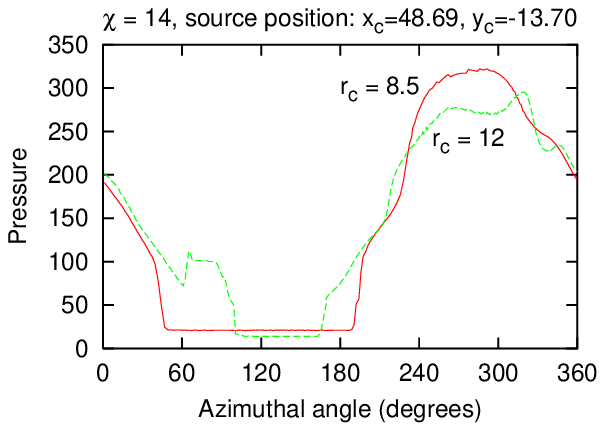,width=8.0cm}
\psfig{figure=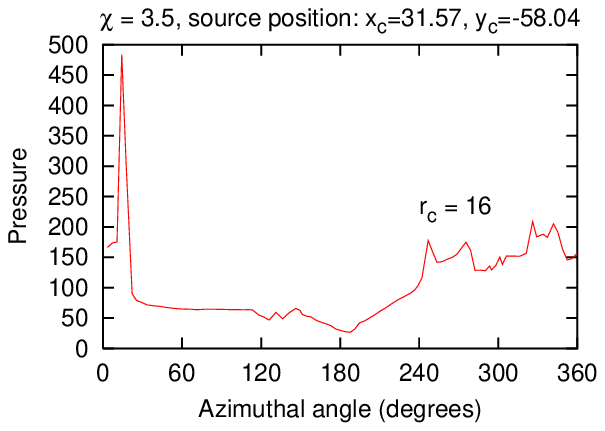,width=8.0cm}
\psfig{figure=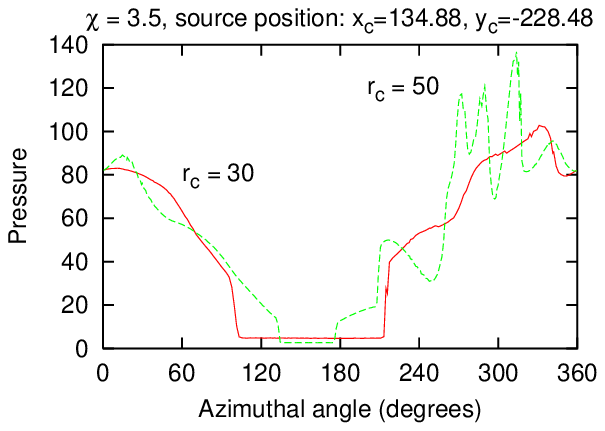,width=8.0cm}
\end{center}
\caption[]{Pressure profiles around some of the mass sources in the
simulations. In each panel the value of $\chi$ and the position of
the mass source are identified. The radius at which the 
pressure profile was taken is noted in each panel. The azimuthal angle
is $0^{\circ}$ on the upstream side of the mass source and increases
clockwise.}
\label{fig:presprofs}
\end{figure*}

The pressure at a given radius from the centre of specific mass
sources is shown in Fig.~\ref{fig:presprofs}. In this figure we have
chosen mass sources with the most uniform pressure
surroundings in the two models considered. We see that the
pressure around a specific mass source tends to be higher when 
$\chi$ is large\footnote{The exception to this is a small region at an
azimuthal angle of $\approx 15^{\circ}$ around the mass source
with position $(x,y)=(31.57,-58.04)$ in the simulation with $\chi=3.5$.}.
When the reverse shock extends to radii exceeding that at which the
pressure profile is obtained, the pressure drops to a low value.
In the simulation with $\chi=3.5$ we typically find that the pressure
is lowest on the downstream side of the mass source, and highest on
the upstream side. However, this \lq memory\rq~of the ambient flow is
reduced as $\chi$ increases - in the top left panel of Fig.~\ref{fig:presprofs}
we see that the maximum pressure occurs on the downstream side 
of the mass source. Another noteable feature is that the 
pressure as a function of azimuthal angle around the mass source 
may be fairly constant when $\chi$ is large. We anticipate that
if the pressure is high and relatively uniform then there will
be a greater probability that the clump could collapse and give
rise to new star formation, than when the pressure is low, or varies
significantly around the clump.

Although we raised the possibility of flickering in 
Section~\ref{sec:intro}, we see little evidence for this in our models.
The horizontal shock in the injected material at $y\approx -10$ 
in the $\chi \approx 3.5$ model (see Fig.~\ref{fig:multiv3} and the
bottom left panel of Fig.~\ref{fig:multiv3b}) was observed to display some
vertical oscillations, but these may be the result of the 
system relaxing to a steady state and we have not evolved the simulation
long enough to eliminate this possibility. We anticipate 
that flickering may be more apparent in a simulation where the mass
injection rate from each source responds to the local flow conditions.

\section{Conclusions}
\label{sec:conclusions}
The results shown in Section~\ref{sec:results} illustrate that a 2D
calculation of mass injection from a cylindrical source can produce
flow features which are similar to those obtained in an axisymmetric
simulation where mass is injected from a spherical source
\citep[\cf][]{FCPDH2002}. For the interaction with a hypersonic wind,
these features include the presence of a bow shock and the fact that
injected material occupies a downstream region with a width
significantly larger than that of the injection region.  For the
interaction with a transonic wind, the tail produced in the
cylindrical case is slightly broader than that produced in the
spherical case, but in both cases the injected material occupies a
region whose width is considerably less than that obtained when the
wind is hypersonic.

When two mass sources are in close proximity, their interaction may
affect the shape and alignment of the tail which forms downstream of
each source. Deviation in the alignment of the tails is
caused by pressure differences either side of each mass source and by
interaction with the reflected bow shock. These induce velocity
components into the flow perpendicular to the upstream velocity of the
ambient wind. A global bow shock envelops the mass sources when they
are close to each other, but increasing their separation leads to an
individual bow shock around each source.

For the transonic case, the curvature of the tails is also
produced by pressure differences. The region between the two tails
acts as a narrow channel, and the curvature of the contact discontinuity
causes the wind material which travels between the tails to be accelerated.
This is accompanied by a pressure drop, and the pressure differences on
either side of the tail force the injected material towards the symmetry
axis. In both the hypersonic and transonic cases, separating the sources 
reduces the degree of the interaction effects.

An interesting result is that in the hypersonic case, tails which show
deviations from alignment with the upstream wind velocity
are pointing away from the
mass sources, whereas in the transonic case the tails end up pointing
towards each other. While this could be a useful way of
determining whether the wind impacting on two mass sources which are
close together is hypersonic or transonic, it is possible that in a 3D
model the wind between the mass sources would not be accelerated as
much, leading to smaller pressure differences on either side of a
tail and less significant curvature. The morphology of the tail 
(broad and \lq stubby\rq, versus long and thin) is instead a much 
simpler way to determine the nature of the interaction. 
 
With multiple mass sources in a hypersonic flow, the ability of the
wind to punch through the space between the sources depends on the
ratio of the mass injection rate from the sources to the mass flux
in the wind, which we have called $\chi$. When $\chi$ is much
greater than unity, the sources are an effective barrier, and allow
very little of the impacting wind to find a path through them. A global
bowshock exists around the sources and the space between the sources is
filled with a high pressure, subsonic flow of injected material. When
$\chi$ is less than or of order unity, the wind is able to force its way
inbetween the mass sources, and for the most part this flow is much 
less pressurized and highly supersonic. We see little evidence for
flickering in our current models.

In future work we will consider 3D simulations, different treatments 
of cooling, and the response of the mass injection rate of 
each source to the local flow conditions. We will also apply these results
to specific objects (\eg, planetary nebulae).

\section*{acknowledgements}
JMP would like to thank PPARC for the funding of a PDRA position and
current funding from the Royal Society. 
This research has made use of NASA's Astrophysics Data System Abstract 
Service.

\bsp

\label{lastpage}

\end{document}